\documentclass[aps, prd, twocolumn, superscriptaddress, floatfix]{revtex4-2}
\usepackage{amsmath}
\usepackage{amssymb}
\usepackage{tikz}
\usepackage{hyperref}
\usepackage{xcolor}

\usetikzlibrary{arrows.meta,calc,decorations,decorations.markings,
  decorations.pathreplacing,decorations.text,fadings,patterns,
  perspective,shadows,shadows.blur,shapes.geometric}
        
\definecolor{linkcolor}{RGB}{0, 0, 255}  
\definecolor{citecolor}{RGB}{0, 128, 0}   
\definecolor{urlcolor}{RGB}{255, 0, 0}    
\hypersetup{
    colorlinks=true,
    linkcolor=linkcolor,
    citecolor=citecolor,
    urlcolor=urlcolor,
    linktoc=all,  
    pdfborder={0 0 0}  
}

\allowdisplaybreaks[4]
\begin{document}
\definecolor{bgcolor}{rgb}{0.95,0.95,0.96}
\definecolor{Navy}{rgb}{0.15,0.30,0.75}
\definecolor{BrickRed}{rgb}{0.80,0.20,0.25}
\definecolor{uc1}{rgb}{0.15,0.35,0.85}
\definecolor{uc2}{rgb}{0.05,0.55,0.55}
\definecolor{uc3}{rgb}{0.20,0.55,0.20}
\definecolor{uc4}{rgb}{0.80,0.50,0.05}
\definecolor{uc5}{rgb}{0.75,0.15,0.20}
\definecolor{platform}{rgb}{0.80,0.84,0.94}
\definecolor{platformlight}{rgb}{0.92,0.94,0.99}
\definecolor{platformedge}{rgb}{0.35,0.45,0.75}
\definecolor{portal}{rgb}{0.42,0.30,0.70}
\definecolor{txt}{rgb}{0.12,0.12,0.15}

\newcommand{\glowline}[3]{
  \draw[#2,line width=#3+2.2pt,opacity=0.18] #1;
  \draw[#2,line width=#3+1.0pt,opacity=0.32] #1;
  \draw[#2,line width=#3,opacity=1] #1;
}

\title{General Construction of Time-Dependent Integrable Chiral Field Theories}

\author{Pradip Kattel}
\email{pradip.kattel@unige.ch}
\affiliation{Department of Quantum Matter Physics, University of Geneva, Quai Ernest-Ansermet 24, 1211 Geneva, Switzerland}

\begin{abstract}
We develop a general procedure for constructing time--dependent integrable chiral field theories from autonomous unitary difference-form $S$-matrices satisfying the Yang--Baxter equation. Spectral parameters are transported along the free right- and left-moving characteristics, defining a map from physical spacetime to spectral space on which the two-body scattering data are evaluated. Requiring spatially homogeneous right-left scattering forces the characteristic map to be affine. The inverse Cayley transform then determines the local contact interaction, while evaluation along the affine spectral trajectory fixes its time dependence. Thus, the nonautonomous interaction is determined by the autonomous scattering data together with chiral kinematics and spatial homogeneity, rather than being introduced independently. The Yang--Baxter equation supplies the factorized many-body transport, and on a spatial circle periodicity leads to quantum Knizhnik--Zamolodchikov (qKZ) equations whose compatibility defines a flat discrete transport in spectral space. Pulling the corresponding spectral-space amplitudes back to physical coordinates gives the time--dependent many-body wavefunctions. Rational $SU(N)$, trigonometric $U_q(\widehat{\mathfrak{sl}}_2)$, and rational $O(N)$ scattering illustrate how the same mechanism generates distinct nonautonomous interactions. The resulting framework gives a geometric route from autonomous factorized scattering data to time--dependent integrable field theories.
\end{abstract}

\maketitle

\section{Introduction}

While the theory of autonomous quantum integrable systems is well established, the systematic understanding of nonautonomous integrability remains much less developed. A few distinct routes have shown that explicit time dependence and quantum integrability can coexist. Driven BCS, Gaudin, Dicke, and multistate Landau--Zener models remain solvable through compatible Hamiltonians and zero-curvature conditions in an auxiliary parameter space~\cite{Sinitsyn2017,Yuzbashyan2018,barik2026higher,patra2015quantum,zabalo2022nonlocality}. Two other directions are particularly relevant to the present work. Extensions of four-dimensional Chern--Simons theory provide a gauge-theoretic route to time-dependent two-dimensional integrable field theories in which the nonautonomous dynamics arises from spacetime-dependent spectral data~\cite{Komatsu2026}. Generalized Bethe Ansatz approaches show that genuinely time-dependent interactions can remain compatible with factorized scattering, with Yang--Baxter consistency and periodicity leading to quantum Knizhnik--Zamolodchikov (qKZ) equations that constrain the admissible time dependence~\cite{PasnooriKondo2025,pasnoori2025exact,PasnooriQKZ2026,PasnooriGN2026,PasnooriRG2026}. The approach developed here is complementary to both: it retains the factorized-scattering and qKZ structure while introducing spacetime dependence through a characteristic map from physical spacetime to spectral space. Our aim is to identify a general mechanism by which autonomous factorized scattering data generate nonautonomous integrable dynamics.

From the viewpoint of factorized scattering, this question suggests a natural inverse problem. An autonomous integrable theory is described at the two-body level by a scattering matrix $S(u)$, whose spectral dependence is constrained by unitarity and Yang--Baxter consistency~\cite{zamolodchikov1979factorized,mussardo1992off}. We ask whether such scattering data can instead serve as the starting point for a local field theory with an explicitly time--dependent Hamiltonian. The problem is then to determine the local interaction together with the relation between the spectral parameter $u$ and physical spacetime.

We address this problem for chiral field theories with linear dispersion. Spectral parameters are transported along the free right- and left-moving characteristics, so their most general spacetime dependence is initially given by arbitrary functions of $x-t$ and $x+t$. We seek spatially homogeneous interactions, for which right-left scattering at a fixed time is independent of the collision position. For a nonconstant difference-form $S$-matrix, this requirement forces both functions to be affine with the same slope and fixes the right-left spectral difference to
\begin{equation}
    u_{LR}(t)=2\kappa t+\beta.
\end{equation}

The local interaction follows directly from the two-body scattering data. For first-order chiral dynamics, integrating the Schr\"odinger equation across a right-left contact gives an exact Cayley relation between the unitary scattering matrix and the Hermitian contact operator~\cite{Schmudgen2012}. Once the normalization of $S(u)$ is fixed, the inverse Cayley transform determines $V(u)$ on its Cayley domain, and evaluation at $u_{LR}(t)=2\kappa t+\beta$ gives the time--dependent Hamiltonian. The scalar phase of $S(u)$ is part of the physical input: although it leaves the Yang--Baxter equation unchanged, it generally changes the contact interaction obtained by the inverse Cayley map.

Geometrically, the characteristic assignment defines a map from physical spacetime to spectral space. Evaluating $S(u)$ and $V(u)$ along this map is their pullback to physical spacetime, so the nonautonomous theory may be viewed as the pullback of an autonomous spectral problem. Spatial homogeneity selects the affine characteristic map above.

In this work, integrability is understood in the factorized-transport sense appropriate to the nonautonomous chiral contact problem. The physical Hamiltonian is time dependent because the two-body scattering data are evaluated along the characteristic map $u=u(x,t)$. At fixed values of the spectral parameters, the two-body matrix $S(u)$ defines an autonomous factorized-scattering problem satisfying the Yang--Baxter equation and the required unitarity conditions.

In the many-body problem, only particles of opposite chirality undergo physical collisions: particles of the same chirality propagate with equal velocity and do not dynamically overtake one another. The Hamiltonian fixes only the right-left scattering. A factorized Bethe Ansatz description nevertheless requires exchange relations between amplitudes in different ordering sectors, including those differing by a same-chirality exchange. The corresponding $RR$ and $LL$ matrices are auxiliary intertwiners of the Bethe Ansatz basis and introduce no additional interaction into the Hamiltonian. Choosing them from the same difference-form solution $S(u)$ places the physical and auxiliary exchanges within a common Yang--Baxter algebra~\cite{andrei1979diagonalization}. On a spatial circle, winding a particle induces a fixed spectral translation and leads to qKZ difference equations~\cite{frenkel1992quantum}. Their compatibility defines a flat discrete transport in spectral space, whose amplitudes are pulled back along the characteristic map to obtain the physical many-body wavefunctions. The reconstruction itself is fixed at the two-body and kinematic levels; representation-dependent methods such as nested Bethe Ansatz or explicit qKZ solutions enter only in constructing particular many-body states~\cite{frenkel1992quantum,Babujian1993,BabujianFlume1994}.

We illustrate the general framework with rational $SU(N)$, trigonometric $U_q(\widehat{\mathfrak{sl}}_2)$, and rational $O(N)$ scattering. Together, these three families demonstrate the applicability of the method across distinct scattering structures and show how affine evolution in spectral space can generate markedly different trajectories of the physical couplings through the inverse Cayley map.

\section{From factorized scattering to a local contact theory}

We begin independently of any particular symmetry algebra. Let $\mathcal{V}$ denote the internal one-particle space and let $\zeta_i$ denote the spectral label carried by particle $i$. We use the ordered-pair convention
\begin{equation}
    u_{ij}:=\zeta_i-\zeta_j, \qquad S_{ij}:=S(u_{ij}).
    \label{eq:orderedSpectralDifference}
\end{equation}
For the physical right-left contact, the relative coordinate is ordered as $r=x_L-x_R$, so the relevant spectral difference is
\begin{equation}
    u_{LR}:=\zeta_L-\zeta_R.
    \label{eq:physicalOrderedDifference}
\end{equation}
The subscript $LR$ records this ordered pair; the interaction is still the physical right-left channel. In the abstract Yang--Baxter equation below, $u,v,w$ denote individual spectral labels, not pairwise differences. We consider a difference-form two-body scattering matrix
\begin{equation}
    S(u)\in \operatorname{End}(\mathcal{V}\otimes\mathcal{V}),
\end{equation}
satisfying the Yang--Baxter equation~\cite{Yang1967,baxter2000partition}
\begin{align}
    &S_{12}(u-v)S_{13}(u-w)S_{23}(v-w)\\
    &=S_{23}(v-w)S_{13}(u-w)S_{12}(u-v).
    \label{eq:YBE}
\end{align}
On the physical spectral line $u\in\mathbb{R}$, we impose both ordinary unitarity and braiding unitarity:
\begin{align}
    S(u)^\dagger S(u)&=\mathbf{1},
    \label{eq:unitarity}\\
    S_{12}(u)S_{21}(-u)&=\mathbf{1}.
    \label{eq:braidingUnitarity}
\end{align}
Here
\begin{equation}
    S_{21}(u)=P S_{12}(u) P,
\end{equation}
with $P$ the permutation operator on $\mathcal{V}\otimes\mathcal{V}$.

For the $P$-invariant scattering matrices considered below, $S_{21}(u)=S_{12}(u)$. Braiding unitarity reduces to $S(u)S(-u)=\mathbf{1}$. We also impose regularity,
\begin{equation}
    S(0)=P.
    \label{eq:regularity}
\end{equation}

Whenever the tensor product $\mathcal{V}\otimes\mathcal{V}$ decomposes into invariant two-particle channels, it is useful to write
\begin{equation}
    \mathcal{V}\otimes\mathcal{V}=\bigoplus_{\alpha}\mathcal{V}_{\alpha}, \quad \mathbf{1} =\sum_{\alpha}\Pi_{\alpha},
\end{equation}
where
\begin{equation}
    \Pi_{\alpha}\Pi_{\beta} = \delta_{\alpha\beta}\Pi_{\alpha}.
\end{equation}
For a multiplicity-free decomposition, the scattering matrix takes the spectral form
\begin{equation}
    S(u) = \sum_{\alpha}s_{\alpha}(u)\Pi_{\alpha}.
    \label{eq:Schannel}
\end{equation}
Unitarity implies $|s_{\alpha}(u)|=1$ for real physical $u$. Equation~\eqref{eq:Schannel} is not required for the construction itself, but it makes the relation between scattering phases and local couplings particularly transparent.

\subsection{Contact reconstruction}

Consider one right mover and one left mover. Their internal two-particle space is
\begin{equation}
    \mathcal H_2=\mathcal V\otimes\mathcal V,
\end{equation}
and we introduce the relative coordinate
\begin{equation}
    r=x_L-x_R.
\end{equation}
For unit chiral velocities, the two-body relative Hamiltonian may be written as
\begin{equation}
    h(u) = 2i\partial_r+2\delta(r)V(u),
    \label{eq:relativeH}
\end{equation}
where
\begin{equation}
    V(u)\in\operatorname{End}(\mathcal H_2), \quad V(u)^\dagger=V(u),
\end{equation}
is the contact operator acting on the internal two-particle space. We use the symmetric contact prescription~\cite{andrei1983solution}
\begin{equation}
    \delta(r)\Psi(r) = \frac{1}{2}\delta(r)\left[\Psi(0^+)+\Psi(0^-)\right].
    \label{eq:symmetriccontact}
\end{equation}

Integrating the Schr\"odinger equation across $r=0$ gives
\begin{equation}
    \left(\mathbf{1}-\frac{i}{2}V\right)\Psi(0^+) = \left(\mathbf{1}+\frac{i}{2}V\right)\Psi(0^-).
\end{equation}
The corresponding scattering matching condition,
\begin{equation}
    \Psi(0^+)=S_V\Psi(0^-),
\end{equation}
is generated by
\begin{equation}
    S_V  = \left(\mathbf{1}-\frac{i}{2}V\right)^{-1}\left(\mathbf{1}+\frac{i}{2}V\right).
    \label{eq:Cayleyforward}
\end{equation}
This is the Cayley relation between the Hermitian contact operator and the unitary scattering matrix. We fix this orientation of the matching condition throughout. Reversing it replaces $S_V$ by $S_V^{-1}=S_{-V}$ and reverses the sign of the corresponding contact operator.

We now impose the prescribed scattering matrix $S(u)$ as the scattering matrix. For real $u$, $S(u)$ is unitary on $\mathcal H_2$. The inverse Cayley transform exists whenever
\begin{equation}
    -1\notin\operatorname{spec}S(u),  \quad \det\left[\mathbf 1+S(u)\right]\neq0,
    \label{eq:CayleyChart}
\end{equation}
where, in the finite-dimensional internal space considered here, the two conditions are equivalent. On this spectral domain, solving the Cayley relation for the contact operator gives
\begin{equation}
    V(u) =2i\left[\mathbf{1}-S(u)\right]\left[\mathbf{1}+S(u) \right]^{-1} \in\operatorname{End}(\mathcal H_2).
    \label{eq:Cayleyinverse}
\end{equation}
The chosen normalization of the scattering matrix sets the contact operator on each Cayley chart. Moreover,
\begin{equation}
    V(u)^\dagger=V(u),
\end{equation}
as follows directly from the unitarity of $S(u)$. The Cayley reconstruction becomes singular precisely when an eigenvalue of $S(u)$ reaches $-1$.

When the channel decomposition of Eq.~\eqref{eq:Schannel} applies, the contact operator is diagonal in the same invariant subspaces,
\begin{equation}
    V(u)  = \sum_{\alpha} v_{\alpha}(u)\Pi_{\alpha},
    \label{eq:Vchannel}
\end{equation}
with
\begin{equation}
    v_{\alpha}(u)  =2i \frac{1-s_{\alpha}(u)} {1+s_{\alpha}(u)}.
    \label{eq:channelCayley}
\end{equation}
Since unitarity gives $|s_\alpha(u)|=1$, each channel eigenvalue may be written as $ s_{\alpha}(u) = e^{i\delta_{\alpha}(u)}$ such that 
\begin{equation}
    v_{\alpha}(u) = 2\tan\frac{\delta_{\alpha}(u)}{2}.
    \label{eq:phasecoupling}
\end{equation}
Thus, the scattering phase in each invariant two-particle channel determines the corresponding real contact coupling.

The scalar normalization of the scattering matrix is part of the inverse-reconstruction data. Let
\begin{equation}
    \widetilde S(u)=\rho(u)S(u),     \qquad |\rho(u)|=1     \quad (u\in\mathbb{R}).
    \label{eq:scalarDress}
\end{equation}
To preserve braiding unitarity and regularity, we require
\begin{equation}
    \rho(u)\rho(-u)=1,     \qquad     \rho(0)=1.
    \label{eq:scalarDressConstraints}
\end{equation}
These conditions leave nontrivial scalar normalization freedom, whose effect on the contact interaction is illustrated explicitly in the trigonometric example below.

\subsection{Second-quantized Hamiltonian and free chiral flow}

Let $\psi_{\chi,a}(x)$ denote chiral fields, with $\chi=+1$ for a right mover and $\chi=-1$ for a left mover, and let $a\in\mathcal V$ be the internal index.  Given $V(u)$ with matrix elements
\begin{equation}
    V(u)^{ab}{}_{cd},
\end{equation}
the corresponding local right-left interaction at fixed spectral coordinate $u$ is
\begin{equation}
    H_{\mathrm{int}}[u] =
    2\int \mathrm{d}x  : \psi^\dagger_{R,a}(x) \psi^\dagger_{L,b}(x)  V(u)^{ab}{}_{cd}  \psi_{L,d}(x) \psi_{R,c}(x) : .
    \label{eq:HintGeneric}
\end{equation} The free Hamiltonian may be written compactly as
\begin{equation}
    H_0  =  -i\sum_{\chi=\pm1}\chi \int \mathrm{d}x \psi_\chi^\dagger(x)\partial_x\psi_\chi(x).
    \label{eq:H0generic}
\end{equation}
For fixed $u$, the autonomous theory is
\begin{equation}
    H[u]=H_0+H_{\mathrm{int}}[u].
    \label{eq:Hfixedu}
\end{equation}

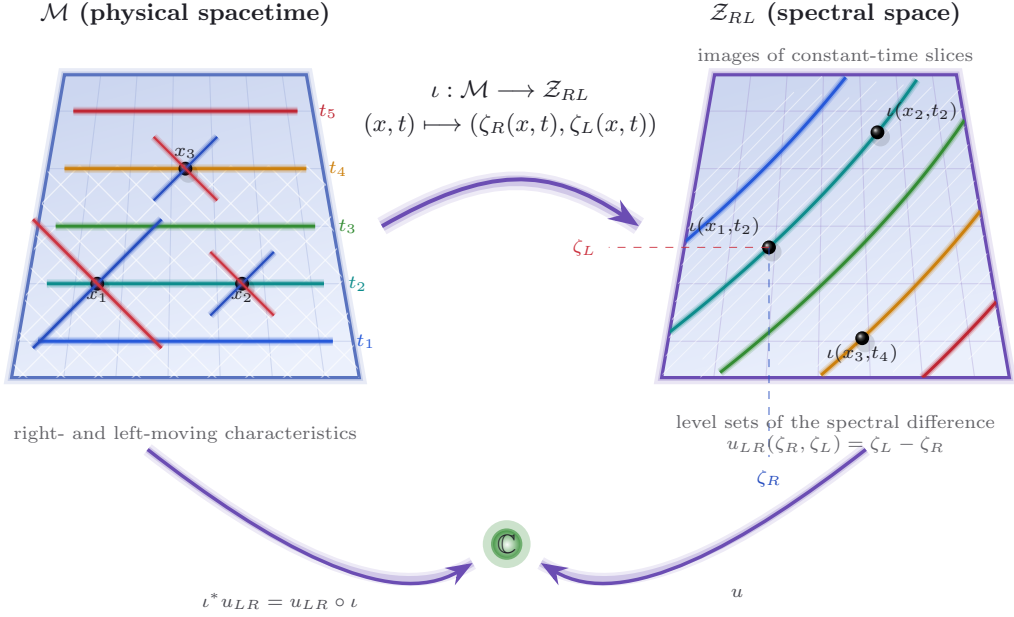
\begin{figure*}
\centering
\begin{tikzpicture}[>=Stealth,every node/.style={font=\small}]


\coordinate (AM) at (0.6,0.2);
\coordinate (BM) at (5.2,0.2);
\coordinate (CM) at (4.5,4.2);
\coordinate (DM) at (1.3,4.2);
\draw[platformedge,line width=5pt,opacity=0.14]
  (AM) -- (BM) -- (CM) -- (DM) -- cycle;
\draw[platformedge,line width=2.5pt,opacity=0.22]
  (AM) -- (BM) -- (CM) -- (DM) -- cycle;
\shade[shading=axis,left color=platform,right color=platformlight,shading angle=15]
  (AM) -- (BM) -- (CM) -- (DM) -- cycle;
\begin{scope}
\clip (AM) -- (BM) -- (CM) -- (DM) -- cycle;
\foreach \tt in {0.12,0.31,0.5,0.69,0.88}{
  \draw[platformedge!70,opacity=0.30,thin] ($(AM)!\tt!(DM)$) -- ($(BM)!\tt!(CM)$);
}
\foreach \ss in {0.15,0.3,0.45,0.6,0.75,0.9}{
  \draw[platformedge!70,opacity=0.30,thin] ($(AM)!\ss!(BM)$) -- ($(DM)!\ss!(CM)$);
}
\end{scope}
\draw[platformedge,line width=1.2pt] (AM) -- (BM) -- (CM) -- (DM) -- cycle;

\coordinate (AZ) at (9.2,0.2);
\coordinate (BZ) at (13.8,0.2);
\coordinate (CZ) at (13.1,4.2);
\coordinate (DZ) at (9.9,4.2);
\draw[portal,line width=5pt,opacity=0.14]
  (AZ) -- (BZ) -- (CZ) -- (DZ) -- cycle;
\draw[portal,line width=2.5pt,opacity=0.22]
  (AZ) -- (BZ) -- (CZ) -- (DZ) -- cycle;
\shade[shading=axis,left color=platform,right color=platformlight,shading angle=15]
  (AZ) -- (BZ) -- (CZ) -- (DZ) -- cycle;
\begin{scope}
\clip (AZ) -- (BZ) -- (CZ) -- (DZ) -- cycle;
\foreach \tt in {0.12,0.31,0.5,0.69,0.88}{
  \draw[portal!70,opacity=0.30,thin] ($(AZ)!\tt!(DZ)$) -- ($(BZ)!\tt!(CZ)$);
}
\foreach \ss in {0.15,0.3,0.45,0.6,0.75,0.9}{
  \draw[portal!70,opacity=0.30,thin] ($(AZ)!\ss!(BZ)$) -- ($(DZ)!\ss!(CZ)$);
}
\end{scope}
\draw[portal,line width=1.2pt] (AZ) -- (BZ) -- (CZ) -- (DZ) -- cycle;

\begin{scope}
\clip (AM) -- (BM) -- (CM) -- (DM) -- cycle;
\foreach \c in {-6,-5.5,...,10}{
  \draw[white,opacity=0.6,line width=0.5pt] (\c-3,-3) -- (\c+3,3);
  \draw[white,opacity=0.6,line width=0.5pt] (\c-3,3) -- (\c+3,-3);
}
\glowline{plot[smooth] coordinates {(0.950,0.680) (1.229,0.680) (1.507,0.680) (1.786,0.680) (2.064,0.680) (2.343,0.680) (2.621,0.680) (2.900,0.680) (3.179,0.680) (3.457,0.680) (3.736,0.680) (4.014,0.680) (4.293,0.680) (4.571,0.680) (4.850,0.680)}}{uc1}{1.1pt}
\glowline{plot[smooth] coordinates {(1.067,1.440) (1.329,1.440) (1.591,1.440) (1.853,1.440) (2.114,1.440) (2.376,1.440) (2.638,1.440) (2.900,1.440) (3.162,1.440) (3.424,1.440) (3.686,1.440) (3.947,1.440) (4.209,1.440) (4.471,1.440) (4.733,1.440)}}{uc2}{1.1pt}
\glowline{plot[smooth] coordinates {(1.184,2.200) (1.429,2.200) (1.674,2.200) (1.919,2.200) (2.165,2.200) (2.410,2.200) (2.655,2.200) (2.900,2.200) (3.145,2.200) (3.390,2.200) (3.635,2.200) (3.881,2.200) (4.126,2.200) (4.371,2.200) (4.616,2.200)}}{uc3}{1.1pt}
\glowline{plot[smooth] coordinates {(1.301,2.960) (1.529,2.960) (1.758,2.960) (1.986,2.960) (2.215,2.960) (2.443,2.960) (2.672,2.960) (2.900,2.960) (3.128,2.960) (3.357,2.960) (3.585,2.960) (3.814,2.960) (4.042,2.960) (4.271,2.960) (4.499,2.960)}}{uc4}{1.1pt}
\glowline{plot[smooth] coordinates {(1.418,3.720) (1.630,3.720) (1.841,3.720) (2.053,3.720) (2.265,3.720) (2.477,3.720) (2.688,3.720) (2.900,3.720) (3.112,3.720) (3.323,3.720) (3.535,3.720) (3.747,3.720) (3.959,3.720) (4.170,3.720) (4.382,3.720)}}{uc5}{1.1pt}
\end{scope}
\node[uc1,anchor=west,font=\scriptsize] at (5.05,0.68) {$t_1$};
\node[uc2,anchor=west,font=\scriptsize] at (4.94,1.44) {$t_2$};
\node[uc3,anchor=west,font=\scriptsize] at (4.81,2.20) {$t_3$};
\node[uc4,anchor=west,font=\scriptsize] at (4.68,2.96) {$t_4$};
\node[uc5,anchor=west,font=\scriptsize] at (4.55,3.72) {$t_5$};

\foreach \p/\lbl/\a in {{1.734,1.440}/x_1/north,{3.650,1.440}/x_2/north,{2.900,2.960}/x_3/south}{
  \fill[black,opacity=0.12] ($(\p)+(0.04,-0.06)$) circle (3.6pt);
  \shade[ball color=black] (\p) circle (2.6pt);
  \node[txt,anchor=\a,font=\scriptsize] at ($(\p)+(0,0.02)$) {$\lbl$};
}
\glowline{(1.734-0.85,1.440-0.85) -- (1.734+0.85,1.440+0.85)}{Navy}{1pt}
\glowline{(1.734-0.85,1.440+0.85) -- (1.734+0.85,1.440-0.85)}{BrickRed}{1pt}
\glowline{(3.650-0.42,1.440-0.42) -- (3.650+0.42,1.440+0.42)}{Navy}{0.8pt}
\glowline{(3.650-0.42,1.440+0.42) -- (3.650+0.42,1.440-0.42)}{BrickRed}{0.8pt}
\glowline{(2.900-0.42,2.960-0.42) -- (2.900+0.42,2.960+0.42)}{Navy}{0.8pt}
\glowline{(2.900-0.42,2.960+0.42) -- (2.900+0.42,2.960-0.42)}{BrickRed}{0.8pt}

\node[txt,font=\bfseries] at (2.9,5.0) {$\mathcal M$ (physical spacetime)};
\node[txt!70,font=\scriptsize] at (2.9,-0.55) {right- and left-moving characteristics};

\begin{scope}
\clip (AZ) -- (BZ) -- (CZ) -- (DZ) -- cycle;
\draw[white,opacity=0.55,very thin] (13.26,0.44) -- (13.68,0.88);
\draw[white,opacity=0.55,very thin] (12.77,0.23) -- (13.64,1.12);
\draw[white,opacity=0.55,very thin] (12.75,0.47) -- (13.60,1.36);
\draw[white,opacity=0.55,very thin] (12.26,0.27) -- (13.55,1.60);
\draw[white,opacity=0.55,very thin] (12.25,0.51) -- (13.51,1.84);
\draw[white,opacity=0.55,very thin] (11.75,0.30) -- (13.47,2.08);
\draw[white,opacity=0.55,very thin] (11.75,0.54) -- (13.43,2.32);
\draw[white,opacity=0.55,very thin] (11.25,0.34) -- (13.39,2.56);
\draw[white,opacity=0.55,very thin] (11.25,0.58) -- (13.34,2.80);
\draw[white,opacity=0.55,very thin] (10.74,0.37) -- (13.30,3.04);
\draw[white,opacity=0.55,very thin] (10.76,0.61) -- (13.26,3.28);
\draw[white,opacity=0.55,very thin] (10.24,0.41) -- (13.22,3.52);
\draw[white,opacity=0.55,very thin] (9.71,0.20) -- (13.18,3.76);
\draw[white,opacity=0.55,very thin] (9.74,0.44) -- (13.13,4.00);
\draw[white,opacity=0.55,very thin] (9.21,0.24) -- (12.80,3.80);
\draw[white,opacity=0.55,very thin] (9.25,0.48) -- (12.77,4.04);
\draw[white,opacity=0.55,very thin] (9.29,0.72) -- (12.42,3.83);
\draw[white,opacity=0.55,very thin] (9.33,0.96) -- (12.40,4.07);
\draw[white,opacity=0.55,very thin] (9.38,1.20) -- (12.05,3.87);
\draw[white,opacity=0.55,very thin] (9.42,1.44) -- (12.04,4.11);
\draw[white,opacity=0.55,very thin] (9.46,1.68) -- (11.68,3.90);
\draw[white,opacity=0.55,very thin] (9.50,1.92) -- (11.68,4.14);
\draw[white,opacity=0.55,very thin] (9.54,2.16) -- (11.32,3.94);
\draw[white,opacity=0.55,very thin] (9.59,2.40) -- (11.32,4.18);
\draw[white,opacity=0.55,very thin] (9.63,2.64) -- (10.95,3.97);
\glowline{plot[smooth] coordinates {(12.650,0.200) (12.825,0.367) (12.994,0.533) (13.159,0.700) (13.319,0.867) (13.475,1.033) (13.625,1.200)}}{uc5}{1.1pt}
\glowline{plot[smooth] coordinates {(11.309,0.233) (11.500,0.400) (11.686,0.567) (11.868,0.733) (12.044,0.900) (12.216,1.067) (12.383,1.233) (12.545,1.400) (12.702,1.567) (12.854,1.733) (13.002,1.900) (13.144,2.067) (13.282,2.233) (13.415,2.400)}}{uc4}{1.1pt}
\glowline{plot[smooth] coordinates {(9.974,0.267) (10.182,0.433) (10.385,0.600) (10.583,0.767) (10.776,0.933) (10.964,1.100) (11.148,1.267) (11.326,1.433) (11.500,1.600) (11.669,1.767) (11.833,1.933) (11.992,2.100) (12.146,2.267) (12.295,2.433) (12.440,2.600) (12.580,2.767) (12.714,2.933) (12.844,3.100) (12.969,3.267) (13.090,3.433) (13.205,3.600)}}{uc3}{1.1pt}
\glowline{plot[smooth] coordinates {(9.305,0.800) (9.515,0.967) (9.719,1.133) (9.919,1.300) (10.114,1.467) (10.305,1.633) (10.490,1.800) (10.670,1.967) (10.846,2.133) (11.017,2.300) (11.183,2.467) (11.344,2.633) (11.500,2.800) (11.651,2.967) (11.798,3.133) (11.939,3.300) (12.076,3.467) (12.208,3.633) (12.335,3.800) (12.457,3.967) (12.574,4.133)}}{uc2}{1.1pt}
\glowline{plot[smooth] coordinates {(9.515,2.000) (9.707,2.167) (9.894,2.333) (10.077,2.500) (10.254,2.667) (10.427,2.833) (10.595,3.000) (10.758,3.167) (10.916,3.333) (11.069,3.500) (11.218,3.667) (11.361,3.833) (11.500,4.000) (11.634,4.167)}}{uc1}{1.1pt}
\end{scope}

\foreach \p/\lbl/\a in {{10.620,1.920}/\iota(x_1{,}t_2)/south east,{12.055,3.440}/\iota(x_2{,}t_2)/south west,{11.853,0.720}/\iota(x_3{,}t_4)/north}{
  \fill[black,opacity=0.12] ($(\p)+(0.04,-0.06)$) circle (3.6pt);
  \shade[ball color=black] (\p) circle (2.6pt);
  \node[txt,anchor=\a,font=\scriptsize] at ($(\p)+(0,0.04)$) {$\lbl$};
}
\draw[Navy,dashed,opacity=0.85] (10.620,1.920) -- (10.620,-0.85);
\node[Navy,anchor=north,font=\scriptsize] at (10.620,-0.9) {$\zeta_R$};
\draw[BrickRed,dashed,opacity=0.85] (10.620,1.920) -- (8.5,1.920);
\node[BrickRed,anchor=east,font=\scriptsize] at (8.45,1.920) {$\zeta_L$};

\node[txt,font=\bfseries] at (11.5,5.0) {$\mathcal Z_{RL}$ (spectral space)};
\node[txt!70,font=\scriptsize,align=center] at (11.5,4.45) {images of constant-time slices};
\node[txt!70,font=\scriptsize,align=center] at (11.5,-0.55) {level sets of the spectral difference\\ $u_{LR}(\zeta_R,\zeta_L)=\zeta_L-\zeta_R$};

\draw[portal,line width=7pt,opacity=0.12]
   (5.5,2.2) .. controls (6.9,3.0) and (7.7,3.0) .. (8.9,2.2);
\draw[portal,line width=4pt,opacity=0.22]
   (5.5,2.2) .. controls (6.9,3.0) and (7.7,3.0) .. (8.9,2.2);
\draw[-{Stealth[length=4mm]},portal,line width=1.5pt]
   (5.5,2.2) .. controls (6.9,3.0) and (7.7,3.0) .. (8.9,2.2);
\node[txt,font=\small,align=center] at (7.2,3.75) {$\iota:\mathcal M\longrightarrow\mathcal Z_{RL}$\\[2pt]$(x,t)\longmapsto(\zeta_R(x,t),\zeta_L(x,t))$};

\coordinate (RR) at (7.15,-2.0);
\draw[portal,line width=5pt,opacity=0.12]
   (2.4,-0.75) .. controls (4.0,-2.0) and (5.6,-2.9) .. (6.7,-2.25);
\draw[portal,line width=2.6pt,opacity=0.22]
   (2.4,-0.75) .. controls (4.0,-2.0) and (5.6,-2.9) .. (6.7,-2.25);
\draw[-{Stealth[length=3.5mm]},portal,line width=1.3pt]
   (2.4,-0.75) .. controls (4.0,-2.0) and (5.6,-2.9) .. (6.7,-2.25);
\node[txt!80,font=\scriptsize] at (4.15,-2.75) {$\iota^*u_{LR}=u_{LR}\circ\iota$};

\draw[portal,line width=5pt,opacity=0.12]
   (11.9,-0.75) .. controls (10.3,-2.0) and (8.7,-2.9) .. (7.6,-2.25);
\draw[portal,line width=2.6pt,opacity=0.22]
   (11.9,-0.75) .. controls (10.3,-2.0) and (8.7,-2.9) .. (7.6,-2.25);
\draw[-{Stealth[length=3.5mm]},portal,line width=1.3pt]
   (11.9,-0.75) .. controls (10.3,-2.0) and (8.7,-2.9) .. (7.6,-2.25);
\node[txt!80,font=\scriptsize] at (10.2,-2.65) {$u$};

\fill[uc3,opacity=0.25] (RR) circle (9pt);
\shade[shading=radial,inner color=white,outer color=uc3,opacity=0.9] (RR) circle (5.5pt);
\draw[uc3!70!white,thick] (RR) circle (5.5pt);
\node[txt,font=\normalsize] at (RR) {$\mathbb C$};

\end{tikzpicture}
\caption{
Characteristic map from physical spacetime to spectral space.
Each spacetime event $(x,t)$ determines the pair of spectral
coordinates $(\zeta_R,\zeta_L)$ associated with the right- and
left-moving characteristics passing through that event. The
colored curves in $\mathcal Z_{RL}$ are the images of constant-time
slices $\mathcal M_{t_i}$ under the characteristic map $\iota$. They
are equivalently level sets of the spectral difference
$u_{LR}(\zeta_R,\zeta_L)=\zeta_L-\zeta_R$.
The pullback $\iota^*u_{LR}=u_{LR}\circ\iota$ gives the spectral coordinate
sampled by the right-left scattering process.
}
\label{fig:pullback}
\end{figure*}

The free chiral motion generated by $H_0$ determines the characteristics along which the spectral parameters will subsequently be transported. Using the canonical equal-time (anti)commutation relations, the Heisenberg equation gives
\begin{equation}
    i\partial_t\psi_\chi(x,t) = [\psi_\chi(x,t),H_0] = -i\chi \partial_x\psi_\chi(x,t),
\end{equation}
or equivalently
\begin{equation}
    D_\chi\psi_\chi=0,  \quad D_\chi:=\partial_t+\chi\partial_x.
    \label{eq:chiralTransportOperator}
\end{equation}
Thus, $D_\chi$ is the characteristic vector field of the free chiral motion. Its characteristic coordinate is
\begin{equation}
    \xi_\chi=x-\chi t, \quad  D_\chi\xi_\chi=0.
    \label{eq:characteristics}
\end{equation}
The right- and left-moving characteristics are $x-t=\mathrm{const}$ and $x+t=\mathrm{const}$, respectively.

\section{Characteristic evaluation and affine spectral evolution}
The preceding analysis gives an autonomous family $H[u]$ labeled by the spectral difference $u$. We now determine how this auxiliary variable acquires physical spacetime dependence. The key point is that the spectral labels are transported with the free chiral motion generated by $H_0$. For chirality $\chi=\pm1$, Eq.~\eqref{eq:chiralTransportOperator} shows that both the field and any label attached to its free trajectory are constant along $D_\chi=\partial_t+\chi\partial_x$. We impose
\begin{equation}
    D_\chi\zeta_\chi(x,t)=0.
    \label{eq:spectralTransportPDE}
\end{equation}
Since the characteristics are $\xi_\chi=x-\chi t$, the general solution is
\begin{equation}
    \zeta_R=f_R(x-t),  \quad \zeta_L=f_L(x+t),
    \label{eq:evaluationmap}
\end{equation}
with otherwise arbitrary differentiable functions $f_R$ and $f_L$. At this stage, $f_R$ and $f_L$ are arbitrary, and their form will be fixed below by requiring spatially homogeneous right-left scattering. The geometric meaning of this characteristic assignment is illustrated
in Fig.~\ref{fig:pullback}.

\subsection{Characteristic pullback from spacetime to spectral space}
The spectral assignment in Eq.~\eqref{eq:evaluationmap} naturally defines a map from physical spacetime to spectral space. Let $\mathcal M$ denote physical spacetime, and let $\mathcal Z_{RL}$ denote the space of right- and left-moving spectral parameters. We define the characteristic map
\begin{equation}
    \iota:\mathcal M\to\mathcal Z_{RL},     \quad     (x,t)\mapsto \left(\zeta_R(x,t), \zeta_L(x,t)\right),
    \label{eq:characteristicMap}
\end{equation}
such that each spacetime event is mapped to the pair of spectral parameters associated with the right- and left-moving characteristics passing through that event as shown in Fig.~\ref{fig:pullback}.

For a difference-form scattering matrix, the relevant spectral coordinate is the function
\begin{equation}
    u_{LR}:\mathcal Z_{RL}\to\mathbb C,     \quad     u_{LR}(\zeta_R,\zeta_L)  =  \zeta_L-\zeta_R,
    \label{eq:spectralDifferenceFunction}
\end{equation}
where the codomain is understood to be restricted to the appropriate spectral domain. Its value on physical spacetime is obtained by composition with the characteristic map,
\begin{equation}
    \iota^*u_{LR} := u_{LR}\circ\iota = f_L(x+t)-f_R(x-t).
    \label{eq:spectralPullback}
\end{equation}
This is the pullback of the spectral coordinate by $\iota$: a function defined on spectral space becomes a function on physical spacetime by evaluation on the spectral parameters assigned by the characteristic map. Here, the superscript $*$ denotes pullback and not Hermitian conjugation.

The above construction applies to the scattering matrix and the contact operator. Define the operator-valued functions on the spectral space 
\begin{align}
    &\mathcal{S}(\zeta_R,\zeta_L) := S(u_{LR})=S(\zeta_L-\zeta_R),\\
    &\mathcal{V}(\zeta_R,\zeta_L):= V(u_{LR})=V(\zeta_L-\zeta_R).
    \label{eq:spectralOperatorFunctions}
\end{align}
Their pullbacks to physical spacetime are
\begin{equation}
    \iota^*\mathcal S =\mathcal S\circ\iota =S\left(\iota^*u_{LR}\right), \quad\iota^*\mathcal V =\mathcal V\circ\iota = V\left(\iota^*u_{LR}\right).
    \label{eq:SVPullback}
\end{equation}
Note that the autonomous functions $S(u)$ and $V(u)$ themselves are not deformed. Their spacetime dependence arises from evaluating them along the characteristic map. Spatial homogeneity will determine the allowed form of this map.

\subsection{Spatial homogeneity and rigidity}
The characteristic assignment in Eq.~\eqref{eq:evaluationmap} generally produces an interaction that depends on both position and time. Here we restrict to nonautonomous Hamiltonians with spatially uniform couplings: at fixed $t$, two right-left collisions occurring at different positions must be governed by the same scattering matrix. This requirement constrains the otherwise arbitrary functions $f_R$ and $f_L$. At a right-left collision, $x_R=x_L=x$, and the spectral difference entering the two-body scattering matrix is
\begin{equation}
    u_{LR}(x,t) =\zeta_L-\zeta_R = f_L(x+t)-f_R(x-t).
    \label{eq:collisionu}
\end{equation}
Spatial homogeneity requires right-left scattering at a fixed time to be independent of the position at which the collision occurs, i.e.,
\begin{equation}
    \partial_x S\left(u_{LR}(x,t)\right)=0.
    \label{eq:homogeneityOperator}
\end{equation}
On a spectral domain where the nonconstant scattering matrix has a nonvanishing spectral derivative, the chain rule gives
\begin{equation}
    \partial_u S(u_{LR}) \partial_x u_{LR}=0.
\end{equation}
Since $\partial_x u_{LR}$ is a scalar and $\partial_u S(u_{LR})\neq0$ on this domain, spatial homogeneity requires
\begin{equation}
    \partial_x u_{LR}=0.
    \label{eq:homogeneity}
\end{equation}
If $\partial_uS$ vanishes only at isolated spectral points, the same condition extends to those points by continuity. Using Eq.~\eqref{eq:collisionu}, we obtain
\begin{equation}
    f_L'(x+t)=f_R'(x-t).
    \label{eq:derivativeRigidity}
\end{equation}

Introduce the light-cone coordinates
\begin{equation}
    p=x+t,  \quad  q=x-t,
\end{equation}
equation~\eqref{eq:derivativeRigidity} becomes
\begin{equation}
    f_L'(p)=f_R'(q).
\end{equation}
Because $p$ and $q$ can be varied independently, this equality can hold for all spacetime points only if both derivatives are equal to the same constant $\kappa$,
\begin{equation}
    f_R'(\xi)=f_L'(\xi)=\kappa.
\end{equation}
Hence, spatial homogeneity fixes the two spectral assignment functions to be affine with a common slope,
\begin{equation}
    f_R(\xi)=\kappa\xi+\nu_R,\quad f_L(\xi)=\kappa\xi+\nu_L.
    \label{eq:affineEvaluation}
\end{equation}
Substituting these functions into Eq.~\eqref{eq:collisionu} gives
\begin{align}
    u_{LR}(x,t) &=\kappa(x+t)+\nu_L -\kappa(x-t)-\nu_R \nonumber\\
    &= 2\kappa t+\beta,  \quad \beta=\nu_L-\nu_R.
    \label{eq:uniformSpectralTranslation}
\end{align}

The collision position $x$ cancels, as required by spatial homogeneity, and every right-left collision at time $t$ samples the same spectral difference $u_{LR}(t)=2\kappa t+\beta$. Collisions at different times sample spectral differences that vary at the constant rate $2\kappa$, even though each individual spectral label remains constant along its own characteristic by Eq.~\eqref{eq:spectralTransportPDE}: the linear drift in $u_{LR}$ comes entirely from the relative motion of the two characteristics, not from any driving imposed on the labels themselves. Free chiral transport together with spatial homogeneity is what fixes the spectral trajectory to be affine.

In the pullback language introduced above, the scattering matrix and contact operator become
\begin{equation}
    \iota^*\mathcal S   =  S(2\kappa t+\beta),  \quad \iota^*\mathcal V  = V(2\kappa t+\beta).
    \label{eq:affineSVPullback}
\end{equation}
The autonomous functions $S(u)$ and $V(u)$ themselves remain unchanged; the characteristic map determines the spectral trajectory on which they are evaluated.

The resulting nonautonomous Hamiltonian is
\begin{equation}
    H(t) = H_0+H_{\mathrm{int}}[2\kappa t+\beta].
    \label{eq:HtimeGeneric}
\end{equation}
The inverse Cayley transform is understood on the corresponding spectral domain,
\begin{equation}
    -1\notin \operatorname{spec}S(2\kappa t+\beta).
\end{equation}
If the affine trajectory reaches a point at which a scattering eigenvalue satisfies $s_\alpha=-1$, the corresponding Cayley coupling $v_\alpha=2\tan(\delta_\alpha/2)$ diverges as the scattering phase passes through $\delta_\alpha=\pi$ modulo $2\pi$. This divergence does not represent a singularity of the underlying scattering problem: the unitary scattering matrix remains well defined, and the matching condition in that channel reduces to $\Psi_\alpha(0+)=-\Psi_\alpha(0-)$. Rather, it marks the breakdown of the particular Cayley parametrization used to represent the contact condition by a finite Hermitian operator $V$. The scattering and matching conditions remain regular at such a point even though the finite-$V$ parametrization becomes singular.

\subsection{Physical scattering and consistency matrices}
The characteristic evaluation also determines the spectral differences for same-chirality pairs.  Since particles of the same chirality propagate with equal velocity, their relative separation is constant, and the explicit time dependence cancels:
\begin{equation}
    \zeta_{\chi,j}-\zeta_{\chi,i} =  \kappa(x_j-x_i), \quad \chi=R,L.
    \label{eq:sameChiralityDifference}
\end{equation}
For an ordered sector with $x_i<x_j$, the exchange convention uses the corresponding first-minus-second argument $\zeta_{\chi,i}-\zeta_{\chi,j}=\kappa(x_i-x_j)$.
This should be distinguished from an opposite-chirality pair, for which the two characteristics intersect and the collision probes the time-dependent spectral difference derived above.  The corresponding scattering matrix is
\begin{equation}
    S_{RL}(t)=S\left(u_{LR}(t)\right)=S(2\kappa t+\beta).
    \label{eq:physicalRL}
\end{equation}

For equal chiralities, there is no corresponding physical collision. Right movers propagate with the same velocity and hence do not dynamically overtake one another, and the same is true for left movers. The Hamiltonian fixes no independent $RR$ or $LL$ scattering amplitudes. This does not, however, remove the need for same-chirality exchange relations in the factorized Bethe Ansatz construction. The many-body wavefunction is represented by amplitudes associated with coordinate-ordering sectors, and consistency of this representation requires relations between amplitudes whose ordering differs by the exchange of two same-chirality particles. The matrices implementing these relations are auxiliary intertwiners of the Bethe Ansatz basis, not additional dynamical scattering matrices. In particular, their introduction adds no $RR$ or $LL$ interaction term to the Hamiltonian.

To obtain a common factorized exchange algebra, we choose these auxiliary matrices from the same difference-form solution $S(u)$,
\begin{equation}
    S_{\chi\chi}^{\mathrm{c}}(i,j)  = S\left(\zeta_{\chi,i}-\zeta_{\chi,j}\right) = S\left(\kappa(x_i-x_j)\right),  \quad \chi=R,L.
    \label{eq:sameChiralityConsistency}
\end{equation}
With this choice, the physical right-left scattering matrix and the auxiliary same-chirality exchange matrices are evaluations of the same autonomous $S(u)$. Hence, the mixed consistency conditions involving $RL$ scattering and $RR$ or $LL$ exchanges are inherited directly from the Yang--Baxter equation~\eqref{eq:YBE}. Thus $S_{RL}$ has a dynamical origin in the contact Hamiltonian, whereas $S_{RR}^{\mathrm{c}}$ and $S_{LL}^{\mathrm{c}}$ have an algebraic role in the Bethe Ansatz basis. They introduce no additional physical scattering process or interaction; they supply only the exchange relations required for a consistent factorized many-body description.

This structure becomes explicit in the many-body problem, to which we now turn.

\section{Many-body transport and qKZ}\label{Many-body-QKZ}
In an autonomous integrable theory, factorized scattering provides the standard route from two-body scattering data to the multiparticle problem~\cite{Yang1967,zamolodchikov1979factorized,andrei1979diagonalization,andrei1983solution,korepin1993quantum}. We now establish this factorized structure for the nonautonomous chiral theory obtained above. Multiparticle amplitudes are related across ordering sectors by pairwise evaluations of $S(\zeta_i-\zeta_j)$, with the Yang--Baxter equation ensuring consistency between different sequences of exchanges. On a spatial circle, periodic transport further induces finite spectral translations and leads to the qKZ difference equations.

\subsection{Transport between ordering sectors}

For $N$ particles, let
\begin{equation}
    \mathcal{C}_Q = \{x_{Q_1}<x_{Q_2}<\cdots<x_{Q_N}\}, \quad   Q\in S_N,
\end{equation}
denote a sector with a fixed ordering of the particle positions. The internal wavefunction in each sector is described by an amplitude
\begin{equation}
    A^Q(\boldsymbol{\zeta}) \in \mathcal{V}^{\otimes N}.
\end{equation}
The amplitudes in neighboring ordering sectors are related by the two-body scattering matrix. We orient an elementary exchange from $Q$ to $Qs_a$ so that
\begin{equation}
    A^{Qs_a}(\boldsymbol{\zeta}) = S_{Q_aQ_{a+1}} \left(\zeta_{Q_a}-\zeta_{Q_{a+1}}\right)A^Q(\boldsymbol{\zeta}),
    \label{eq:sectorTransport}
\end{equation}
where $s_a$ exchanges the particles in positions $a$ and $a+1$. For a physical right-left contact this convention agrees with the matching orientation fixed in Eq.~\eqref{eq:Cayleyforward}: taking $Q=(L,R)$ gives $r=x_L-x_R<0$, while $Qs_a=(R,L)$ gives $r>0$, and hence Eq.~\eqref{eq:sectorTransport} reads $\Psi(0^+)=S(\zeta_L-\zeta_R)\Psi(0^-)$. The reverse exchange is given by the inverse relation. Repeated application of Eq.~\eqref{eq:sectorTransport} determines the amplitudes in all ordering sectors from a reference amplitude. For three or more particles, a given ordering can be reached through different sequences of pairwise exchanges. The Yang--Baxter equation ensures that all such sequences give the same amplitude.

These relations retain the standard autonomous factorized form in spectral space. Their nonautonomous realization in physical spacetime is obtained by evaluating the spectral parameters on the characteristic map derived above.

\subsection{Periodic transport and qKZ}

We now compactify physical space to a circle,
\begin{equation}
    x\sim x+L.
\end{equation}
Winding particle $j$ once around the circle shifts its coordinate by $x_j\to x_j+L$. By Eq.~\eqref{eq:affineEvaluation}, this physical winding induces a fixed translation of its spectral parameter,
\begin{equation}
    \zeta_j(x_j+L,t) = \zeta_j(x_j,t)+p, \quad p=\kappa L.
    \label{eq:qKZShift}
\end{equation}

During the winding, particle $j$ is successively scattered through the remaining particles. The corresponding ordered product of two-body scattering matrices defines the transport operator $K_j(\boldsymbol{\zeta};p)$. For a conventional reference ordering,
\begin{align}
K_j(\boldsymbol{\zeta};p)=&S_{j,j-1}\left(\zeta_j-\zeta_{j-1}+p\right)\cdots S_{j1} \left(\zeta_j-\zeta_1+p\right)
\nonumber\\
&\times S_{jN}\left(\zeta_j-\zeta_N\right)\cdots
S_{j,j+1}\left(\zeta_j-\zeta_{j+1}\right).
    \label{eq:qKZTransport}
\end{align}

Since the winding simultaneously shifts $\zeta_j\to\zeta_j+p$, the periodic transport relation takes the form of a spectral difference equation,
\begin{equation}
    \Phi(\boldsymbol{\zeta}+p e_j) = K_j(\boldsymbol{\zeta};p) \Phi(\boldsymbol{\zeta}), \quad     j=1,\ldots,N.
    \label{eq:qKZ}
\end{equation}
These are the qKZ transport equations associated with the difference-form scattering matrix. Their compatibility requires
\begin{equation}
    K_i(\boldsymbol{\zeta}+p e_j;p) K_j(\boldsymbol{\zeta};p) = K_j(\boldsymbol{\zeta}+p e_i;p) K_i(\boldsymbol{\zeta};p).
    \label{eq:qKZCompatibility}
\end{equation}
Equation~\eqref{eq:qKZCompatibility} states that successive translations of two spectral parameters give the same amplitude independently of their order. The operators $K_j$ define a flat discrete transport in spectral space.

\subsection{Characteristic pullback to physical wavefunctions}

The characteristic map extends naturally to the $N$-particle configuration space. Let $\mathcal M_N$ denote configuration spacetime, with coordinates $(\boldsymbol{x},t)$, and let $\mathcal Z_N$ denote the corresponding spectral space, with coordinates $\boldsymbol{\zeta}=(\zeta_1,\ldots,\zeta_N)$. We define 
\begin{equation}
    \iota_N:\mathcal M_N\to\mathcal Z_N, \quad (\boldsymbol{x},t) \mapsto \left( \zeta_1(x_1,t),\ldots,\zeta_N(x_N,t)   \right),
    \label{eq:manyBodyCharacteristicMap}
\end{equation}
where
\begin{equation}
    \zeta_j(x_j,t)  =\kappa(x_j-\chi_j t)+\nu_{\chi_j}, \quad  \chi_j=\pm1.
    \label{eq:physicalEvaluation}
\end{equation}

Let $\Phi_Q:\mathcal Z_N\to\mathcal V^{\otimes N}$ be a spectral-space amplitude satisfying the factorized transport and qKZ relations above. Its  pullback along $\iota_N$ defines the corresponding time-dependent many-body wavefunction,
\begin{equation}
    \Psi_Q =\iota_N^*\Phi_Q := \Phi_Q\circ\iota_N,
    \label{eq:manyBodyPullback}
\end{equation}
or explicitly,
\begin{equation}
    \Psi_Q(\boldsymbol{x},t) = \Phi_Q\left(\zeta_1(x_1,t),\ldots,\zeta_N(x_N,t) \right).
    \label{eq:qKZPhysicalEvaluation}
\end{equation}
Notice that the physical spacetime dependence of the many-body wavefunction is carried entirely by the characteristic spectral parameters $\zeta_j(x_j,t)$.

The relation between physical winding and spectral translation can now be stated directly at the level of the characteristic map. Define
\begin{equation}
    T_j(\boldsymbol{x},t) := (\boldsymbol{x}+Le_j,t),  \quad \tau_j(\boldsymbol{\zeta})  := \boldsymbol{\zeta}+p e_j.
\end{equation}
Equation~\eqref{eq:qKZShift} is then equivalently the intertwining relation
\begin{equation}
    \iota_N\circ T_j = \tau_j\circ\iota_N.
    \label{eq:intertwining}
\end{equation}
Hence, the characteristic map sends a physical winding of particle $j$ to the corresponding qKZ translation in spectral space.

Pulling back Eq.~\eqref{eq:qKZ} along $\iota_N$ gives
\begin{equation}
    \Psi_Q(\boldsymbol{x}+Le_j,t) = K_j \left( \iota_N(\boldsymbol{x},t);p \right)
    \Psi_Q(\boldsymbol{x},t).
    \label{eq:physicalQKZTransport}
\end{equation}
It is convenient to define the corresponding physical transport operator
\begin{equation}
    \mathcal K_j(\boldsymbol{x},t) := K_j \left( \iota_N(\boldsymbol{x},t);p \right).
    \label{eq:physicalTransportOperator}
\end{equation}
The qKZ compatibility condition~\eqref{eq:qKZCompatibility} then pulls
back to
\begin{equation}
    \mathcal K_i(\boldsymbol{x}+Le_j,t) \mathcal K_j(\boldsymbol{x},t) = \mathcal K_j(\boldsymbol{x}+Le_i,t) \mathcal K_i(\boldsymbol{x},t).
    \label{eq:physicalFlatness}
\end{equation}
Thus the flat discrete transport in spectral space induces a flat discrete transport on the configuration space through the characteristic map. Here ``flat'' refers to the consistency of these discrete winding transports and does not imply an independent continuous zero-curvature connection in physical spacetime.

For a specified scattering algebra, solutions of Eq.~\eqref{eq:qKZ} may be constructed using the corresponding off-shell Bethe Ansatz machinery ~\cite{frenkel1992quantum,Babujian1993,BabujianFlume1994,PasnooriQKZ2026}. Higher-rank theories lead to nested qKZ systems; this representation- dependent step is separate from the two-body reconstruction above.

\section{Schr\"odinger evolution from the characteristic evaluation}

We now verify that the time-dependent many-body wavefunction obtained by characteristic evaluation solves the Schr\"odinger problem for the reconstructed contact Hamiltonian. The contact interaction acts only when a right- and left-moving particle meet; between such contacts, the wavefunction obeys the free chiral evolution.

Let particle $j$ have chirality $\chi_j=\pm1$ and spectral parameter
\begin{equation}
    \zeta_j(x_j,t) = \kappa(x_j-\chi_j t)+\nu_{\chi_j}.
    \label{eq:tdseEvaluation}
\end{equation}
For a spectral-space amplitude $\Phi_Q(\boldsymbol{\zeta})$, its characteristic evaluation in the ordering sector $Q$ is
\begin{equation}
    \Psi_Q(\boldsymbol{x},t) =\Phi_Q\left(\zeta_1(x_1,t),\ldots,\zeta_N(x_N,t)\right).
    \label{eq:tdsePullback}
\end{equation}
By the chain rule,
\begin{equation}
    \partial_t\Psi_Q =\sum_{j=1}^{N}\frac{\partial\Phi_Q}{\partial\zeta_j}\partial_t\zeta_j,
\end{equation}
and
\begin{equation}
    \partial_{x_j}\Psi_Q = \frac{\partial\Phi_Q}{\partial\zeta_j} \partial_{x_j}\zeta_j.
\end{equation}
Since
\begin{equation}
    \partial_t\zeta_j=-\chi_j\kappa, \quad \partial_{x_j}\zeta_j=\kappa,
\end{equation}
we obtain
\begin{equation}
    \left(\partial_t +\sum_{j=1}^{N}\chi_j\partial_{x_j}\right)\Psi_Q(\boldsymbol{x},t) = 0.
    \label{eq:freeCharacteristicEquation}
\end{equation}

The free first-quantized $N$-particle Hamiltonian is
\begin{equation}
    H_0^{(N)} = -i\sum_{j=1}^{N} \chi_j\partial_{x_j},
    \label{eq:firstQuantizedH0}
\end{equation}
so Eq.~\eqref{eq:freeCharacteristicEquation} is precisely 
\begin{equation}
    i\partial_t\Psi_Q = H_0^{(N)}\Psi_Q
\end{equation}
within each ordering sector. This is the expected bulk equation for a contact interaction: the interaction enters at the boundaries $x_i=x_j$ rather than in the interior of the sectors.

At a right-left contact, the amplitudes on the two sides of the boundary are related by the Cayley matching condition, with the orientation specified in Eq.~\eqref{eq:sectorTransport}, evaluated at the collision spectral coordinate $u_{LR}(t)=2\kappa t+\beta$. Thus, the free evolution within the ordering sectors and the time-dependent matching at their right-left boundaries together give the Schr\"odinger problem generated by $H(t)$. Successive right-left contacts in the many-body problem may occur at different physical times. Their chronological order determines the sequence in which the corresponding sector matchings are encountered, but does not introduce a separate time-dependent deformation of the scattering algebra. Since each $\zeta_j$ is transported unchanged along its characteristic, every contact samples the ordered pairwise spectral difference $\zeta_i-\zeta_j$ appearing in the spectral-space transport relations of Sec.~\ref{Many-body-QKZ}. The consistency of different admissible sequences of such matchings is the Yang--Baxter consistency already established in spectral space; the explicit time dependence enters only through the characteristic evaluation of the spectral parameters at the physical contacts.

Therefore, a spectral-space solution $\Phi_Q(\boldsymbol{\zeta})$ satisfying the factorized transport relations determines the full time-dependent interacting many-body wavefunction through
\begin{equation}
    \Psi_Q(\boldsymbol{x},t) = \Phi_Q\left(\kappa(x_1-\chi_1t)+\nu_{\chi_1},\ldots,\kappa(x_N-\chi_Nt)+\nu_{\chi_N}\right).
    \label{eq:fullTimeDependentWavefunction}
\end{equation}
On a spatial circle, these wavefunctions also satisfy the qKZ winding relations derived above. 

We now apply the general construction to several distinct families of factorized scattering matrices. Each normalized unitary difference-form $S$-matrix satisfying the assumptions above determines, on its admissible Cayley domain, through characteristic evaluation a corresponding family of nonautonomous integrable chiral field theories. The examples below illustrate the range of interactions generated by this procedure, from rational higher-rank models to anisotropic trigonometric and orthogonal families. 

\section{Rational \texorpdfstring{$SU(N)$}{SU(N)} scattering}

We begin with the rational fundamental $SU(N)$ scattering matrix
\begin{equation}
    S_N(u) = \frac{ u \mathbf{1}  +  i\eta P }{ u+i\eta },
    \label{eq:SUNS}
\end{equation}
where $P$ permutes the two fundamental internal spaces, $\eta\in\mathbb{R}$, and satisfies
\begin{equation}
    P^2=\mathbf{1}.
\end{equation}
The normalization in Eq.~\eqref{eq:SUNS} makes the symmetric channel trivial. Introducing
\begin{equation}
    \Pi_{+}=\frac{\mathbf{1}+P}{2}, \quad \Pi_{-}=\frac{\mathbf{1}-P}{2},
\end{equation}
one finds
\begin{equation}
    S_N(u) = \Pi_{+} +\frac{u-i\eta}{u+i\eta}\Pi_{-}.
\end{equation}
For $u\neq0$ (the Cayley chart is singular at $u=0$, where the antisymmetric-channel eigenvalue is $-1$), the channel Cayley transform Eq.~\eqref{eq:channelCayley} gives
\begin{equation}
    v_{+}(u)=0, \quad v_{-}(u)=-\frac{2\eta}{u},
\end{equation}
or equivalently
\begin{equation}
    V_N(u) = \frac{\eta}{u} \left( P-\mathbf{1} \right).
    \label{eq:SUNV}
\end{equation}

Evaluating this kernel on the affine collision trajectory gives the driven $SU(N)$ theory.  In second-quantized form,
\begin{equation}
\begin{split}
    &H_{SU(N)}(t) = -i\int \mathrm{d}x \psi^\dagger_{R,a}\partial_x\psi_{R,a} + i\int \mathrm{d}x \psi^\dagger_{L,a}\partial_x\psi_{L,a} \\
    &\quad+ 2g(t)\int \mathrm{d}x :\psi^\dagger_{R,a}\psi^\dagger_{L,b}\left( P-\mathbf{1}\right)^{ab}{}_{cd}
    \psi_{L,d}\psi_{R,c} : ,
\end{split}
    \label{eq:SUNHamiltonian}
\end{equation}
where
\begin{equation}
    g(t) = \frac{\eta}{2\kappa t+\beta}.
    \label{eq:SUNg}
\end{equation}
Equation~\eqref{eq:SUNHamiltonian} is the contact Hamiltonian fixed by the normalization in Eq.~\eqref{eq:SUNS}; matching it to a particular Gross--Neveu or Thirring convention in the literature is only a matter of Fierz rearrangement.

\section{Trigonometric \texorpdfstring{$U_q(\widehat{\mathfrak{sl}}_2)$}{Uq(sl2)} scattering}

We next consider the fundamental trigonometric
$U_q(\widehat{\mathfrak{sl}}_2)$ solution. Let $0<\gamma<\pi$ be the
anisotropy parameter and define
\begin{equation}
R_\gamma(u)=
\begin{pmatrix}
\sinh(u+i\gamma) & 0 & 0 & 0\\
0 & \sinh u & i\sin\gamma & 0\\
0 & i\sin\gamma & \sinh u & 0\\
0 & 0 & 0 & \sinh(u+i\gamma)
\end{pmatrix}.
\end{equation}

We use the regular braiding-unitary normalization
\begin{equation}
S_\gamma(u)= \frac{R_\gamma(u)}{\sinh(u+i\gamma)}.
\label{eq:TrigS}
\end{equation}

For real $u$, this matrix satisfies
\begin{equation}
S_\gamma(u)^\dagger S_\gamma(u)=\mathbf{1}_4,
\quad S_\gamma(u)S_\gamma(-u)=\mathbf{1}_4,
\quad S_\gamma(0)=P.
\label{eq:TrigUnitarity}
\end{equation}

The two parallel-spin states have eigenvalue
\begin{equation}
s_{\parallel}(u)=1.
\label{eq:TrigParallelEigenvalue}
\end{equation}

The symmetric and antisymmetric states in the zero-magnetization sector
have eigenvalues
\begin{equation}
s_{\pm}(u) = \frac{\sinh u\pm i\sin\gamma}
{\sinh(u+i\gamma)}.
\label{eq:TrigMixedEigenvalues}
\end{equation}

Applying the inverse Cayley transform gives
\begin{align}
V_\gamma(u) &=
-\tan\frac{\gamma}{2}\coth u \left(\mathbf{1}-\sigma^z\otimes\sigma^z\right)\nonumber\\
&~+\frac{\tan(\gamma/2)}{\sinh u}\left(\sigma^x\otimes\sigma^x+\sigma^y\otimes\sigma^y\right).
\label{eq:TrigV}
\end{align}

The corresponding channel eigenvalues are
\begin{equation}
v_{\parallel}(u)=0,
\label{eq:TrigVParallel}
\end{equation}
and
\begin{equation}
v_{+}(u) =-2\tan\frac{\gamma}{2}\tanh\frac{u}{2},\qquad
v_{-}(u)=-2\tan\frac{\gamma}{2}\coth\frac{u}{2}.
\label{eq:TrigVMixed}
\end{equation}

The Cayley representation is singular at $u=0$, where the
antisymmetric-channel eigenvalue of $S_\gamma(u)$ equals $-1$.
The scattering matrix itself remains finite and regular there because
$S_\gamma(0)=P$.

The characteristic construction evaluates the spectral coordinate on the affine collision trajectory
\begin{equation}
u(t)=2\kappa t+\beta.
\label{eq:TrigTrajectory}
\end{equation}

Hence,
\begin{align}
V_\gamma(t)&=-\tan\frac{\gamma}{2}\coth(2\kappa t+\beta)\left(\mathbf{1}-\sigma^z\otimes\sigma^z\right)\nonumber\\
&~+\frac{\tan(\gamma/2)}{\sinh(2\kappa t+\beta)}\left(\sigma^x\otimes\sigma^x+
\sigma^y\otimes\sigma^y\right).
\label{eq:TrigVTime}
\end{align}

The corresponding Hamiltonian is
\begin{equation}
\begin{split}
H_{\mathrm{trig}}(t)={}&-i\int \mathrm{d}x\psi^\dagger_{R,a}\partial_x\psi_{R,a}
+i\int \mathrm{d}x\psi^\dagger_{L,a}\partial_x\psi_{L,a}\\
&+2\int \mathrm{d}x :\psi^\dagger_{R,a}\psi^\dagger_{L,b} V_\gamma(2\kappa t+\beta)^{ab}{}_{cd} \psi_{L,d}\psi_{R,c}: .
\end{split}
\label{eq:TrigHamiltonian}
\end{equation}

To illustrate the physical effect of the residual scalar normalization, choose the admissible factor
\begin{equation}
    \rho_a(u)=e^{iau},    \qquad a\in\mathbb{R},
\end{equation}
which satisfies the constraints in Eq.~\eqref{eq:scalarDressConstraints}. The dressed trigonometric matrix is
\begin{equation}
    \widetilde S_{\gamma,a}(u)  =  e^{iau}S_\gamma(u).
\end{equation}

For the parallel-spin channel, the undressed eigenvalue is $s_{\parallel}(u)=1$, so that
\begin{equation}
    v_{\parallel}(u)=0.
\end{equation}
After the admissible scalar dressing,
\begin{equation}
    \widetilde v_{\parallel}(u) = 2i\frac{1-e^{iau}}{1+e^{iau}} =    2\tan\frac{au}{2}.
\end{equation}
Thus, the same Yang--Baxter, unitary, braiding-unitary, and regular scattering structure can produce either a zero or a nonzero contact coupling, depending on the allowed scalar normalization.

\section{Orthogonal \texorpdfstring{$O(N)$}{O(N)} scattering}

We next consider the rational $O(N)$ scattering family, which provides a distinct class of nonautonomous integrable theories generated by the inverse construction. For $N>2$, the vector representation of $O(N)$ admits three invariant two-particle tensor structures. In addition to the identity and permutation operators, one has the trace, or contraction, operator
\begin{equation}
    K^{ab}{}_{cd}  = \delta^{ab}\delta_{cd}.
    \label{eq:Kdefinition}
\end{equation}
For the corresponding many-particle internal space $\mathcal V^{\otimes m}$, the nearest-neighbor contraction operators $K_i=K_{i,i+1}$ generate a representation of the Temperley--Lieb algebra with loop parameter $N$~\cite{temperley2004relations}. For the two-particle operators relevant here,
\begin{equation}
    P^2=\mathbf{1},  \quad PK=KP=K, \quad K^2=NK.
    \label{eq:ONalgebra}
\end{equation}
The two-particle space decomposes into antisymmetric, symmetric-traceless, and singlet channels,
\begin{equation}
    \mathcal{V}\otimes\mathcal{V}  =  \mathcal{V}_{A}  \oplus  \mathcal{V}_{T}  \oplus
    \mathcal{V}_{0},
\end{equation}
with projectors
\begin{equation}
    \Pi_A =  \frac{\mathbf{1}-P}{2},
    \label{eq:ONPiA}
\end{equation}
and
\begin{equation}
    \Pi_T     =   \frac{\mathbf{1}+P}{2} - \frac{K}{N},  \qquad  \Pi_0  =  \frac{K}{N}.
    \label{eq:ONPiT0}
\end{equation}

For the rational $O(N)$ scattering matrix ~\cite{zamolodchikov1978relativistic,babujian2016bethe}, we adopt the parametrization of Andrei and Destri~\cite{AndreiDestri1984},
\begin{equation}
    S_{O(N)}(u)  = \frac{1}{iu+2}  \left[  iu \mathbf{1}   + 2P  - \frac{2iu}{iu+N-2}K \right].
    \label{eq:ONS}
\end{equation}
For real $u$, $S_{O(N)}(u)$ is unitary and satisfies the difference-form Yang--Baxter equation. Its channel eigenvalues follow
directly from Eqs.~\eqref{eq:ONalgebra}--\eqref{eq:ONPiT0}.

In the antisymmetric channel,
\begin{equation}
    s_A(u)  =   \frac{iu-2}{iu+2},
    \label{eq:ONsA}
\end{equation}
while in the symmetric-traceless channel,
\begin{equation}
    s_T(u)=1.
    \label{eq:ONsT}
\end{equation}
In the singlet channel, using $P\Pi_0=\Pi_0$ and $K\Pi_0=N\Pi_0$, one finds
\begin{equation}
    s_0(u)  = \frac{ (iu-2)(iu-N+2) }{  (iu+2)(iu+N-2)  }.
    \label{eq:ONs0}
\end{equation}
For real $u$, all three eigenvalues have unit modulus, and hence
\begin{equation}
    S_{O(N)}(u)^\dagger S_{O(N)}(u)  =   \mathbf{1}.
\end{equation}

Applying the inverse Cayley transform channel by channel gives
\begin{equation}
    v_A(u)  =  2i\frac{1-s_A(u)}{1+s_A(u)}  =\frac{4}{u},  \qquad  v_T(u)=0,
    \label{eq:ONvAT}
\end{equation}
and
\begin{equation}
    v_0(u) = 2i\frac{1-s_0(u)}{1+s_0(u)}  = \frac{2Nu}{u^2-2(N-2)}.
    \label{eq:ONv0}
\end{equation}
Thus, on the corresponding Cayley domain,
\begin{equation}
    V_{O(N)}(u)  =  \frac{4}{u}\Pi_A  + \frac{2Nu}{u^2-2(N-2)}\Pi_0.
    \label{eq:ONVchannels}
\end{equation}
Using Eqs.~\eqref{eq:ONPiA} and \eqref{eq:ONPiT0}, this becomes
\begin{equation}
    V_{O(N)}(u)  = \frac{2}{u} \left( \mathbf{1}-P \right)  + \frac{2u}{u^2-2(N-2)}K.
    \label{eq:ONV}
\end{equation}
The spectral dependence is carried by two distinct scalar channel couplings, while the operator structure is fixed by the orthogonal invariant tensors.

For real $u$, the finite-$V$ Cayley parametrization becomes
singular at
\[
u=0,\qquad u^2=2(N-2),
\]
where the antisymmetric and singlet scattering eigenvalues, respectively, reach $-1$. The scattering matrix itself remains finite and unitary at these points; the singularity occurs in its representation by the contact operator $V_{O(N)}(u)$. Away from these points, $V_{O(N)}(u)$ is finite and Hermitian.

Evaluating Eq.~\eqref{eq:ONV} on the affine collision trajectory
\begin{equation}
    u(t)=2\kappa t+\beta
\end{equation}
gives the corresponding nonautonomous $O(N)$ theory,
\begin{equation}
\begin{split}
    &H_{O(N)}(t) =  -i\int \mathrm{d}x   \psi^\dagger_{R,a}\partial_x\psi_{R,a}
    + i\int \mathrm{d}x   \psi^\dagger_{L,a}\partial_x\psi_{L,a}   \\
    &\quad+ 2\int \mathrm{d}x  : \psi^\dagger_{R,a} \psi^\dagger_{L,b}   V_{O(N)}(2\kappa t+\beta)^{ab}{}_{cd} \psi_{L,d}\psi_{R,c} : .
\end{split}
    \label{eq:ONHamiltonian}
\end{equation}
The corresponding eigenvalues of the contact operator are
\begin{equation}
    v_A(t)   =  \frac{4}{2\kappa t+\beta},~~  v_T(t)=0,  ~~  v_0(t)
    =  \frac{ 2N(2\kappa t+\beta) }{ (2\kappa t+\beta)^2-2(N-2)  }.
    \label{eq:ONcouplings}
\end{equation}
The antisymmetric and singlet sectors acquire distinct time-dependent contact interactions, both set by the single affine spectral trajectory.

\section{Conclusion and outlook}

In this work, we have presented a general construction of nonautonomous integrable chiral field theories from autonomous factorized scattering data. The central result is that the local interaction and its admissible time dependence are not independent. The former is fixed by the normalized scattering matrix through the inverse Cayley map, while the latter is fixed by chiral characteristics together with spatial homogeneity. The same framework extends to the factorized many-body problem and determines the full time-dependent interacting wavefunctions through characteristic evaluation. The rational $SU(N)$, trigonometric $U_q(\widehat{\mathfrak{sl}}_2)$, and rational $O(N)$ examples show that affine evolution in spectral space can produce markedly different trajectories in coupling space. They also expose the role of scalar normalization in the inverse reconstruction. Scattering matrices that differ only by a scalar phase, and hence have the same Yang--Baxter structure, can yield different local contact interactions through the inverse Cayley map.

The restriction to spatially homogeneous interactions can be relaxed at the level of the characteristic map. Without imposing homogeneity, the spectral assignments retain the general form $\zeta_R=f_R(x-t)$ and $\zeta_L=f_L(x+t)$, and the Cayley reconstruction produces the spatially inhomogeneous interaction $V(x,t)=V \left(f_L(x+t)-f_R(x-t)\right)$. The characteristic pullback and local contact matching remain well defined in this setting, while the periodic problem becomes richer. For nonaffine $f_R$ and $f_L$, winding around the spatial circle no longer induces a constant spectral shift, and the fixed-step qKZ equations derived here must be replaced by a more general transport problem. Understanding the corresponding consistency conditions may provide a route to integrable theories with genuinely spacetime-dependent couplings. A second extension is to relax the assumption of linear chiral dispersion. Fixed right- and left-moving velocities make the present characteristic map particularly simple, with spectral labels transported along $x-\chi t=\mathrm{const}$. For a nonlinear dispersion, the characteristic velocity generally depends on momentum or rapidity, and the spectral assignment would acquire the corresponding kinematic dependence. The underlying principle may nevertheless persist. Spectral data may be transported along the physical trajectories of the excitations, with the autonomous scattering data evaluated on the resulting map. Establishing whether such momentum-dependent characteristic maps remain compatible with factorized scattering and a suitable generalization of qKZ transport could extend this approach to a broader class of nonautonomous integrable field theories.

The geometric formulation presented here also suggests a natural treatment of interactions localized along specific worldlines or boundaries. A localized impurity provides one such setting and defines a defect worldline along which integrable defect scattering data can be evaluated~\cite{delfino1994scattering,konik1999purely}. We conjecture that the present construction extends to such systems by supplementing the bulk $S$-matrix with the appropriate defect transmission and reflection matrices, imposing the corresponding factorization relations, and evaluating the resulting scattering data along the defect worldline to obtain nonautonomous impurity couplings. This may provide a route to time-dependent integrable impurity and Kondo-type models. Open boundaries suggest a parallel possibility, with the bulk Yang--Baxter structure supplemented by a boundary reflection matrix satisfying the reflection equation~\cite{sklyanin1988boundary,ghoshal1993boundary} and evaluated along the boundary to generate nonautonomous integrable boundary interactions.

Recent time-dependent Bethe Ansatz and qKZ results have identified integrable coupling trajectories with renormalization-group flows~\cite{PasnooriRG2026,PasnooriGN2026}. In the framework developed here, the inverse Cayley map determines the couplings as functions of the spectral coordinate $u$, while the characteristic map fixes $u(t)=2\kappa t+\beta$. If the resulting curves in coupling space coincide with RG trajectories of the corresponding autonomous theories, the physical time evolution generated here follows the same trajectories, with $t$ providing an affine parametrization of RG time. Determining whether this relation holds for the scattering families considered here would clarify how generally integrable time evolution can be related to renormalization-group flow.

Four-dimensional Chern--Simons theory provides a complementary perspective on the role of spectral data in integrable field theories. The gauge-theoretic approach of Costello, Witten, and Yamazaki gives a geometric origin to Yang--Baxter structures and two-dimensional integrable field theories~\cite{costello2017gauge,costello2018gauge,costello2019gauge}, while recent extensions generate nonautonomous theories through spacetime-dependent spectral data~\cite{Komatsu2026}. In the present approach, by contrast, the autonomous spectral data remain fixed, and spacetime dependence enters through the characteristic map on which they are evaluated. A gauge-theoretic interpretation of the characteristic map could clarify the relation between these two mechanisms.
 \section{Acknowledgments}
 We thank Parameshwar Pasnoori for an insightful conversation that motivated us to investigate the general principles underlying time--dependent Bethe Ansatz solvability. This work was supported by the Swiss National Science Foundation under Division II (Grant No.~200020-219400).

\bibliographystyle{unsrt}
\bibliography{ref}

@article{AndreiDestri1984,
  title={Dynamical symmetry breaking and fractionization in a new integrable model},
  author={Andrei, N and Destri, C},
  journal={Nuclear Physics B},
  volume={231},
  number={3},
  pages={445--480},
  year={1984},
  publisher={Elsevier}
}

@article{Sinitsyn2017,
  title={Integrable time-dependent quantum Hamiltonians},
  author={Sinitsyn, Nikolai A and Yuzbashyan, Emil A and Chernyak, Vladimir Y and Patra, Aniket and Sun, Chen},
  journal={Physical review letters},
  volume={120},
  number={19},
  pages={190402},
  year={2018},
  publisher={APS}
}

@article{Yuzbashyan2018,
  title={Integrable time-dependent Hamiltonians, solvable Landau--Zener models and Gaudin magnets},
  author={Yuzbashyan, Emil A},
  journal={Annals of Physics},
  volume={392},
  pages={323--339},
  year={2018},
  publisher={Elsevier}
}

@article{barik2026higher,
  title={Higher-spin Richardson-Gaudin model with time-dependent coupling: Exact dynamics},
  author={Barik, Suvendu and Bakker, Lieuwe and Gritsev, Vladimir and Min{\'a}{\v{r}}, Ji{\v{r}}{\'\i} and Yuzbashyan, Emil A},
  journal={Physical Review B},
  volume={113},
  number={19},
  pages={195147},
  year={2026},
  publisher={APS}
}

@article{PasnooriKondo2025,
  title={Integrability of the Kondo model with time-dependent interaction strength},
  author={Pasnoori, Parameshwar R},
  journal={Physical Review B},
  volume={112},
  number={6},
  pages={L060409},
  year={2025},
  publisher={APS}
}

@article{PasnooriQKZ2026,
  title={Quantum Knizhnik-Zamolodchikov equations and integrability of quantum field theories with time-dependent interaction strength},
  author={Pasnoori, Parameshwar R},
  journal={Physical Review B},
  volume={113},
  number={9},
  pages={094514},
  year={2026},
  publisher={APS}
}

@article{PasnooriRG2026,
  title={Quantum integrability of Hamiltonians with time-dependent interaction strengths and the renormalization group flow},
  author={Pasnoori, Parameshwar R},
  journal={Physical Review B},
  volume={113},
  number={20},
  pages={L201405},
  year={2026},
  publisher={APS}
}

@article{PasnooriGN2026,
  title={Time-Dependent Dynamical Dimensional Transmutation in the $ SU (2) $ Gross-Neveu Model with Time-Dependent Interaction Strength},
  author={Pasnoori, Parameshwar R},
  journal={arXiv preprint arXiv:2605.05111},
  year={2026}
}

@article{Komatsu2026,
  title={Time-Dependent Integrability from Gauge Theory, I},
  author={Komatsu, Shota and Sakamoto, Jun-ichi and Wallberg, Anders and Yamazaki, Masahito},
  journal={arXiv preprint arXiv:2607.02648},
  year={2026}
}

@article{Yang1967,
  title={Some exact results for the many-body problem in one dimension with repulsive delta-function interaction},
  author={Yang, Chen-Ning},
  journal={Physical Review Letters},
  volume={19},
  number={23},
  pages={1312},
  year={1967},
  publisher={APS}
}

@article{Babujian1993,
  title={Off-shell Bethe ansatz equations and N-point correlators in the SU (2) WZNW theory},
  author={Babujian, Hrachia M},
  journal={Journal of Physics A: Mathematical and General},
  volume={26},
  number={23},
  pages={6981--6990},
  year={1993}
}

@article{BabujianFlume1994,
  title={Off-shell Bethe ansatz equation for Gaudin magnets and solutions of Knizhnik-Zamolodchikov equations},
  author={Babujian, Hrachia M and Flume, Rainald},
  journal={arXiv preprint hep-th/9310110},
  year={1993}
}

@article{mussardo1992off,
  title={Off-critical statistical models: factorized scattering theories and bootstrap program},
  author={Mussardo, Giuseppe},
  journal={Physics Reports},
  volume={218},
  number={5-6},
  pages={215--379},
  year={1992},
  publisher={Elsevier}
}

@article{zamolodchikov1979factorized,
  title={Factorized S-matrices in two dimensions as the exact solutions of certain relativistic quantum field theory models},
  author={Zamolodchikov, Alexander B and Zamolodchikov, Alexey B},
  journal={Annals of physics},
  volume={120},
  number={2},
  pages={253--291},
  year={1979},
  publisher={Elsevier}
}

@Inbook{Schmudgen2012,
author="Schm{\"u}dgen, Konrad",
title="Self-adjoint Extensions: Cayley Transform and Krein Transform",
bookTitle="Unbounded Self-adjoint Operators on Hilbert Space",
year="2012",
publisher="Springer Netherlands",
address="Dordrecht",
pages="283--306",
isbn="978-94-007-4753-1",
doi="10.1007/978-94-007-4753-1_13",
url="https://doi.org/10.1007/978-94-007-4753-1_13"
}

@article{frenkel1992quantum,
  title={Quantum affine algebras and holonomic difference equations},
  author={Frenkel, Igor B and Reshetikhin, N Yu},
  journal={Communications in mathematical physics},
  volume={146},
  number={1},
  pages={1--60},
  year={1992},
  publisher={Springer}
}

@article{andrei1979diagonalization,
  title={Diagonalization of the chiral-invariant Gross-Neveu Hamiltonian},
  author={Andrei, N and Lowenstein, Jo H},
  journal={Physical review letters},
  volume={43},
  number={23},
  pages={1698},
  year={1979},
  publisher={APS}
}

@article{baxter2000partition,
  title={Partition function of the eight-vertex lattice model},
  author={Baxter, Rodney J},
  journal={Annals of Physics},
  volume={281},
  number={1-2},
  pages={187--222},
  year={2000},
  publisher={Elsevier}
}

@article{andrei1983solution,
  title={Solution of the Kondo problem},
  author={Andrei, Natan and Furuya, K and Lowenstein, JH},
  journal={Reviews of modern physics},
  volume={55},
  number={2},
  pages={331},
  year={1983},
  publisher={APS}
}

@incollection{temperley2004relations,
  title={Relations between the ‘percolation’and ‘colouring’problem and other graph-theoretical problems associated with regular planar lattices: some exact results for the ‘percolation’problem},
  author={Temperley, Harold NV and Lieb, Elliott H},
  booktitle={Condensed Matter Physics and Exactly Soluble Models: Selecta of Elliott H. Lieb},
  pages={475--504},
  year={2004},
  publisher={Springer}
}

@article{zamolodchikov1978relativistic,
  title={Relativistic factorized S-matrix in two dimensions having O (N) isotopic symmetry},
  author={Zamolodchikov, Alexander B and Zamolodchikov, Alexey B},
  journal={Nuclear Physics B},
  volume={133},
  number={3},
  pages={525--535},
  year={1978},
  publisher={Elsevier}
}

@article{babujian2016bethe,
  title={Bethe Ansatz and exact form factors of the O (N) Gross Neveu-model},
  author={Babujian, Hrachya M and Foerster, Angela and Karowski, Michael},
  journal={Journal of High Energy Physics},
  volume={2016},
  number={2},
  pages={1--33},
  year={2016},
  publisher={Springer}
}

@article{costello2017gauge,
  title={Gauge theory and integrability, I},
  author={Costello, Kevin and Witten, Edward and Yamazaki, Masahito},
  journal={arXiv preprint arXiv:1709.09993},
  year={2017}
}

@article{costello2018gauge,
  title={Gauge theory and integrability, II},
  author={Costello, Kevin and Witten, Edward and Yamazaki, Masahito},
  journal={arXiv preprint arXiv:1802.01579},
  year={2018}
}

@article{costello2019gauge,
  title={Gauge theory and integrability, III},
  author={Costello, Kevin and Yamazaki, Masahito},
  journal={arXiv preprint arXiv:1908.02289},
  year={2019}
}

@book{korepin1993quantum,
  title={Quantum inverse scattering method and correlation functions},
  author={Korepin, Vladimir E and Bogoliubov, Nikolai Mikhailovich and Izergin, Anatoli G},
  volume={9},
  year={1993},
  publisher={Cambridge university press Cambridge}
}

@article{delfino1994scattering,
  title={Scattering theory and correlation functions in statistical models with a line of defect},
  author={Delfino, Gesualdo and Mussardo, Giuseppe and Simonetti, P},
  journal={Nuclear Physics B},
  volume={432},
  number={3},
  pages={518--550},
  year={1994},
  publisher={Elsevier}
}

@article{konik1999purely,
  title={Purely transmitting defect field theories},
  author={Konik, Robert and LeClair, Andre},
  journal={Nuclear Physics B},
  volume={538},
  number={3},
  pages={587--611},
  year={1999},
  publisher={Elsevier}
}

@article{ghoshal1993boundary,
  title={Boundary S-matrix and boundary state in two-dimensional integrable quantum field theory},
  author={Ghoshal, Subir and Zamolodchikov, Alexander},
  journal={arXiv preprint hep-th/9306002},
  year={1993}
}

@article{sklyanin1988boundary,
  title={Boundary conditions for integrable quantum systems},
  author={Sklyanin, Evgeni K},
  journal={Journal of Physics A: Mathematical and General},
  volume={21},
  number={10},
  pages={2375--2389},
  year={1988}
}

@article{pasnoori2025exact,
  title={Exact many-body wavefunction of the Kondo model with time-dependent interaction strength},
  author={Pasnoori, Parameshwar R and Yuzbashyan, Emil and others},
  journal={arXiv preprint arXiv:2509.05640},
  year={2025}
}

@article{patra2015quantum,
  title={Quantum integrability in the multistate Landau--Zener problem},
  author={Patra, Aniket and Yuzbashyan, Emil A},
  journal={Journal of Physics A: Mathematical and Theoretical},
  volume={48},
  number={24},
  pages={245303},
  year={2015},
  publisher={IOP Publishing}
}

@article{zabalo2022nonlocality,
  title={Nonlocality as the source of purely quantum dynamics of BCS superconductors},
  author={Zabalo, Aidan and Wu, Ang-Kun and Pixley, JH and Yuzbashyan, Emil A},
  journal={Physical Review B},
  volume={106},
  number={10},
  pages={104513},
  year={2022},
  publisher={APS}
}

\end{document}